\documentclass{PoS}

\usepackage{epsfig}

\PoS{PoS(LAT2005)197}

\title{Two-colour Lattice QCD with dynamical fermions at non-zero density
  versus Matrix Models}

\ShortTitle{$N_c=2$ Lattice QCD with dynamical fermions at $\mu\neq0$ 
vs. Matrix Models}

\author{\speaker{Gernot Akemann}\\
        Department of Mathematical Sciences\\
        Brunel University West London\\
        Uxbridge UB8 3PH, United Kingdom\\
        E-mail: \email{gernot.akemann@brunel.ac.uk}}

\author{Elmar Bittner\\
        Institut f\"ur Theoretische Physik\\ 
        Universit\"at Leipzig\\ 
        Augustusplatz 10/11\\
        D-04109 Leipzig, Germany\\
        E-mail: \email{elmar.bittner@itp.uni-leipzig.de}}

\abstract{We provide first evidence that Matrix Models describe the low lying
  complex Dirac eigenvalues in a theory with dynamical fermions at non-zero
  density. Lattice data for gauge group $SU(2)$ with staggered fermions are
  compared to detailed analytical results from Matrix Models in the
  corresponding symmetry class, the complex chiral Symplectic Ensemble. They
  confirm the predicted dependence on chemical potential, quark mass and
  volume. 
}

\FullConference{XXIIIrd International Symposium on Lattice Field Theory\\
		 25-30 July 2005\\
		 Trinity College, Dublin, Ireland}

\begin{document}

\section{Introduction}\label{intro}

In the past few years many new analytical results have emerged from
Matrix Models (MM) for the Dirac operator spectrum with non-vanishing chemical
potential $\mu\neq0$. 
There are 3 different possible chiral symmetry breaking patterns
\cite{Jac3fold} and in the MM picture they differ crucially in the way
the eigenvalues are depleted from the imaginary axis through $\mu\neq0$
\cite{HOV}. 
Today we have detailed predictions for 
microscopic Dirac spectra in 2 of these symmetry classes,
both quenched and unquenched: QCD and the adjoined representation, where the
latter is replaced by $SU(2)$ colour in the fundamental when using staggered
fermions as we will do.
After first approximate results for quenched QCD \cite{A02} the exact quenched
density was derived in \cite{SV} and related to chiral Perturbation Theory in
the epsilon regime ($\epsilon\chi$PT). 
Unquenched partition functions were computed in \cite{AFV} and related to
$\epsilon\chi$PT, and finally fully 
unquenched Dirac spectra for QCD became available \cite{J,AOSV}.
Results for the adjoint (or $SU(2)$ staggered) class followed very recently
\cite{A05} and we confront its unquenched predictions including
dependence on $\mu$ and quark mass $m$ with Lattice data here. 

Not all the above MM results have been compared to the Lattice so far,
precisely due to the sign problem in unquenched QCD. 
Up to now only quenched simulations at $\mu\neq0$ using staggered fermions 
have been successfully described: for
QCD \cite{AW} (see also \cite{James} in these proceedings) and for $SU(2)$
\cite{ABLMP}, which we will extend here.
Previous comparisons \cite{BLMP} were lacking analytic predictions at the time,
they could only be done in the bulk of the spectrum \cite{MPW} 
for MM without chiral symmetry. Very recently it 
has been shown at $\mu=0$ that the previous topology-blindness
of staggered fermions can be cured by improvement, as reviewed in
\cite{Edouardo}. 

In the last years different ways of attacking the sign problem in 
QCD were developed: multi-parameter
reweighting, Taylor-expansion and imaginary $\mu$
(see \cite{Karsch} for a review and references).
However, none of these have been applied so far to the region of
$\epsilon\chi$PT where a comparison to MM is expected to hold.
We purse a different avenue here, choosing an $SU(2)$ gauge
theory without sign problem where
dynamical simulations can be performed in a standard way \cite{HKLM}. 
This permits us to
extend previous MM comparisons \cite{BMW,AK} to $\mu\neq0$. 
It is of principle interest to
test the validity of all 
MM predictions for complex Dirac spectra on the Lattice
including dynamical fermions.

\section{Predictions from Matrix Models with $\mu\neq0$}\label{MM}

In this section we briefly introduce the relevant MM used and give its
results for the spectral density. 
For more details and references 
we refer to \cite{A05}. The MM partition function is given by
\begin{equation}
{\cal Z}_N^{(2N_f)}(\{m_f\};\mu) \equiv
 \int\! d\Phi  d\Psi 
\exp\left[-N\mbox{Tr} (\Phi^\dag \Phi\ +\ \Psi^\dagger\Psi)\right]
\prod_{f=1}^{N_f} 
\det\left( \begin{array}{cc}
m_f\mbox{\bf 1} & i \Phi + \mu \Psi \\
i \Phi^{\dagger} + \mu \Psi^{\dagger} & m_f\mbox{\bf 1}
\end{array} \right) \!,
\label{Z2MM}
\end{equation}
where \mbox{\bf 1} is the quaternion unity element. The 
two rectangular $(N+\nu)\times N$
matrices, $\Phi$ and $\Psi$, contain quaternion real elements and replace the
off-diagonal blocks of the Dirac operator,
averaged with a Gaussian
matrix weight instead of the gauge action. Here we model the
$\mu\gamma_0$ part with a random matrix of the same symmetry as the kinetic
part, assuming it is non-diagonal in 
matrix space (as in \cite{J} for QCD). If universality holds this choice of
basis should not matter, compared to $\Psi$ replaced by unity as \cite{HOV}.
This assertion is true in the QCD
symmetry class, see \cite{SV,AFV} vs. \cite{J,AOSV}. 
The size of the Dirac matrix $2N\sim V$
relates to the volume, and we have chosen rectangular matrices to be in the
fixed sector of $\nu\geq0$ 
zero eigenvalues or topological charge. Because of using
unimproved staggered fermions we set $\nu=0$ in our comparison to the data
later. 

After transforming the linear combinations $i \Phi^{(\dagger)} + \mu
\Psi^{(\dagger)}$ to be triangular we can change to complex eigenvalues
$z_{k=1,\ldots,N}$ (rotated to lie on the real axis for $\mu=0$). 
All their spectral correlation functions can be computed using
skew orthogonal Laguerre polynomials in the complex plane \cite{A05}.  
In the large-$N$ limit the complex eigenvalues, masses {\it and } chemical
potential have to be rescaled 
\begin{equation}
\sqrt{2}\ Nz \ \equiv \xi, \ \ \sqrt{2}\ Nm\ \equiv\ \eta,\ \ 
\mbox{and } \ \ \lim_{N\to\infty,\ \mu\to0}2N\mu^2\ \equiv \ \alpha^2 \ .
\label{weak}
\end{equation}
This limit is called weakly non-hermitian as $\mu^2$
is rescaled with the volume. The same scaling is found in the static
part of the chiral Lagrangian \cite{Kogut} dominating at $\epsilon\chi$PT.
While the latter contains {\it two}
parameters, the chiral condensate $\Sigma$ and the decay constant $F_\pi$, the
MM contains only {\it one}, the variance fixed to be $\Sigma=\sqrt{2}$ here. 
Thus the constant multiplying $\alpha$ is not know and we are left with 1 free
fit parameter, in contrast to the parameter-free MM prediction at
$\mu=0$\footnote{This observation was missed in \cite{AW,ABLMP}.}.

The unquenched microscopic spectral density, normalised to 
$\Im m(\xi)\delta'(\Im m(\xi))$ for $\alpha\to0$,
\begin{eqnarray}
\rho_{weak}^{(4)}(\xi;\eta) &=& \Delta\rho_{weak}^{(4)}(\xi;\eta) 
\ +\ \frac{1}{32\alpha^4}
(\xi^{\ast\,2}-\xi^2)\ |\xi|^2\ 
K_{2\nu}\left(\frac{|\xi|^2}{2\alpha^2}\right)
\exp\left[+\frac{1}{4\alpha^2}(\xi^2+\xi^{*\,2})\right]
\label{rhoKweakmass}\\
&\times&\int_0^1 ds \int_0^1 \frac{dt}{\sqrt{t}}\mbox{e}^{-2s(1+t)\alpha^2}
\left(J_{2\nu}(2\sqrt{st}\ \xi)J_{2\nu}(2\sqrt{s}\ \xi^\ast)
\ -\ J_{2\nu}(2\sqrt{s}\ \xi)J_{2\nu}(2\sqrt{st}\ \xi^\ast)\right),
\nonumber
\end{eqnarray}
splits into the quenched part 
and a correction term $\Delta\rho_{weak}^{(4)}(\xi;\eta)$ depending on 
the mass $\eta$:
\begin{eqnarray}
&&\Delta\rho_{weak}^{(4)}(\xi;\eta) \ =\  
\frac{1}{32\alpha^4}
(\xi^{\ast\,2}-\xi^2)\ |\xi|^2\ 
K_{2\nu}\left(\frac{|\xi|^2}{2\alpha^2}\right)
\exp\left[+\frac{1}{4\alpha^2}(\xi^2+\xi^{*\,2})\right]
\label{deltarhoweak}\\
&&\times\left\{ 
\left( 
\int_0^1\!\!ds\int_0^1dt \sqrt{\frac{s}{t}}\ \mbox{e}^{-2s(1+t)\alpha^2}
( J_{2\nu}(2\sqrt{st}\ \xi)I_{2\nu+1}(2\sqrt{s}\ \eta)
- \sqrt{t}I_{2\nu+1}(2\sqrt{st}\ \eta) J_{2\nu}(2\sqrt{s}\ \xi))
\right)\right.\nonumber\\
&&\times\left.\left( 
\int_0^1\!\!ds\int_0^1\frac{dt}{\sqrt{t}}\ \mbox{e}^{-2s(1+t)\alpha^2}
( J_{2\nu}(2\sqrt{st}\, \xi^*)I_{2\nu}(2\sqrt{s}\, \eta)
- I_{2\nu}(2\sqrt{st}\, \eta) J_{2\nu}(2\sqrt{s}\, \xi^*))\!\!
\right)\!
-(\xi\leftrightarrow\xi^*)\!\right\}\nonumber\\
&&\times
\left[\int_0^1\!\!ds\int_0^1dt \sqrt{\frac{s}{t}}\ \mbox{e}^{-2s(1+t)\alpha^2}
\left( \sqrt{t} I_{2\nu+1}(2\sqrt{st}\ \eta)I_{2\nu}(2\sqrt{s}\ \eta)
- I_{2\nu}(2\sqrt{st}\ \eta)I_{2\nu+1}(2\sqrt{s}\ \eta)\right) \right]^{-1}\!.
\nonumber
\end{eqnarray}
We only give the result
for $N_f=2$ staggered flavours of equal mass here, see
\cite{A05} for more flavours and higher correlation functions, as
well as \cite{BMW} for matching MM and staggered flavours. 
\begin{figure}[-h]
\centering
\includegraphics[width=17pc]{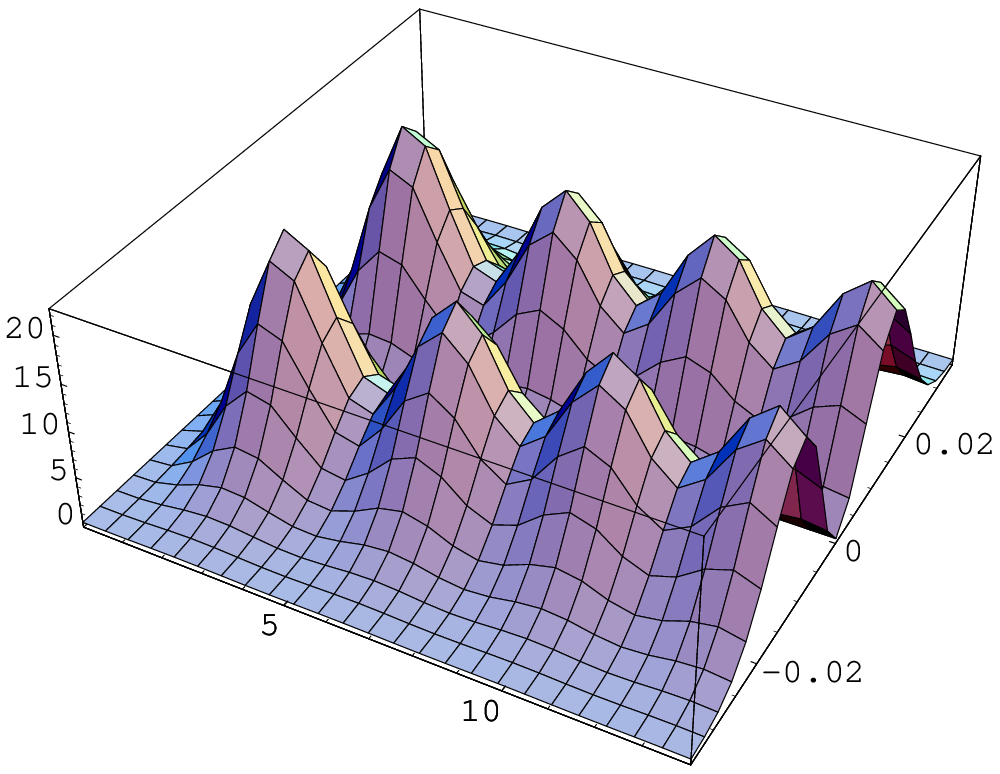}
\includegraphics[width=17pc]{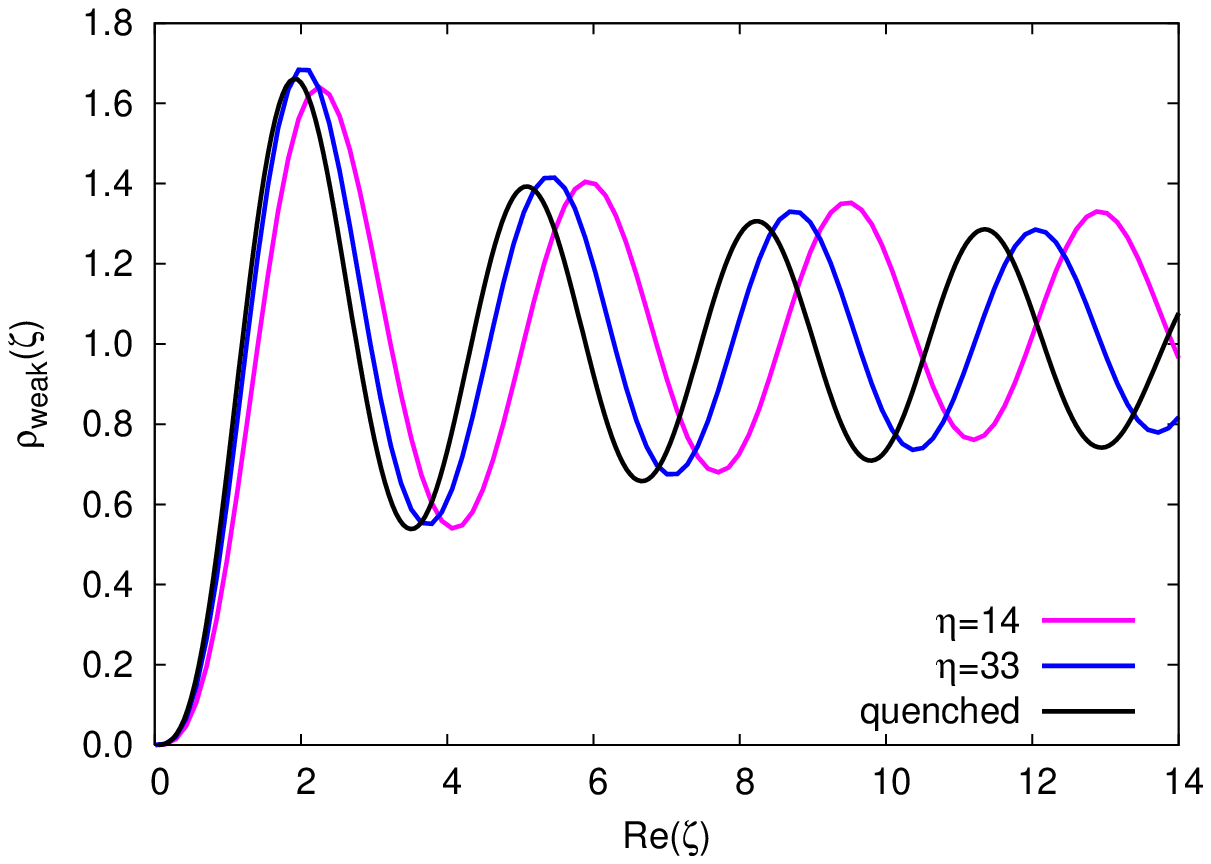}
\put(-370,10){\scriptsize $\Re e(\xi)$}
\put(-230,40){\scriptsize $\Im m(\xi)$}
\put(-410,120){\scriptsize $\rho_{weak}^{(4)}(\xi)$}
\caption{The $N_f=2$ flavour spectral density at $\alpha=0.012$ with masses
  $\eta=8.74$ and $\nu=0$ (left) and a cut through the density normalised to
  unity  
  for different mass values compared to quenched (right).} 
\label{rhoweakmassplot}
\end{figure}
Fig. \ref{rhoweakmassplot} shows this density in the complex plane at 
weak non-Hermiticity, as we shall compare to data below. The eigenvalues
are repelled from the real axis for $\mu\neq0$,
being a distinct feature of this symmetry class (compare to \cite{AW} for
QCD). We have observed this
repulsion in the data for  $\mu$ as small as $10^{-6}$. The individual
eigenvalues located at the maxima previously are now split into a double peak
in the complex plane. 

Increasing the mass $\eta\to\infty$ moves the density to the left, bringing it
back to the quenched expression, eq. (\ref{rhoKweakmass}) at
$\Delta\rho_{weak}^{(4)}(\xi;\eta)=0$.  
Decreasing $\eta\to0$ pushes
the eigenvalues further away from the origin, approaching the quenched density
at $\nu=2$ approximately. Increasing $\alpha$ rapidly washes out the
oscillations and leads to the formation of a plateau. The limit
$\alpha\to\infty$ takes us to the MM at
strong non-Hermiticity, with $\mu$ {\it unscaled} (see \cite{A05}). A
comparison to quenched Lattice data in this regime was given previously in
\cite{ABLMP}, our unquenched results will be reported elsewhere.

\section{Lattice data with dynamical fermions at $\mu\neq0$}\label{Lat}

Our data were generated for gauge group $SU(2)$ with coupling 
$\beta=4/g^2=1.3$ and $N_f=2$ staggered flavours, using the code of 
\cite{HKLM}. In this setup the fermion
determinant remains real (as it does in the MM, see \cite{A05}) and standard
Monte Carlo applies. We have studied two different Volumes $6^4$ and $8^4$ 
for various values of $\mu=10^{-6}-0.4$ and values of the quark masses
$ma=0.025-20$. Because the eigenvalues lie in the complex plane of the order
of 5-10k configurations are needed.
In order to have a window where a MM description applies for these 
small lattices we have to go to relatively strong coupling (see \cite{James}
for a discussion of this issue in QCD).

To compare with the prediction (\ref{rhoKweakmass}) 
we have taken cuts through the
density: along the maxima parallel to the real axis, 
Fig. \ref{rhoweakmassplot} right,  
and perpendicular to that
over the first maximum pair.
The effect of dynamical fermions is most clearly seen in the shift in the
first cut where we choose the values
$\eta=14$ (blue) and  $\eta=33$ (pink) to be used in our data below,
compared to quenched (black).
Being very costly we did not to go to smaller $\eta$.

The parameters $\eta$ and $\alpha$ were obtained as follows from our
data with input $ma$ and $\mu$. 
First, we determined the rescaling of the masses and eigenvalues by
measuring the mean 
level-spacing $d\sim1/\rho(0)$, using the Banks-Casher relation
from $\mu=0$: $\pi\rho(0)=\Sigma V$ where $\rho(0)$ is the mean spectral
density. Due to $\mu\ll1$ the 
geometric distance between eigenvalues
agrees within errors with the distance obtained by a projection onto the real
axis. This provides us with the rescaling of the eigenvalues and masses,
\begin{equation}
za\pi /d \ \equiv \xi \ \ \mbox{and} \ \ 
ma\pi/d \ \equiv \eta \ .
\end{equation}
At the same time the spacing $d$ contains the volume factor for the 
rescaling of $\mu^2$: $\alpha^2= C\mu^2\pi/d$. The constant $C\sim
F_\pi^2/\Sigma$  of order 1 
is obtained by fitting the data to the cut parallel to the
imaginary axis on the first maximum. 
Since the surface under this curve is finite
we can fit to its integral function,
being independent of the choice of histogram
widths in imaginary direction. 
This also fixes the normalisation. For illustration purposes 
we show the cut and not its integral.
\begin{figure}[-h]
\centering
\includegraphics[width=17pc]{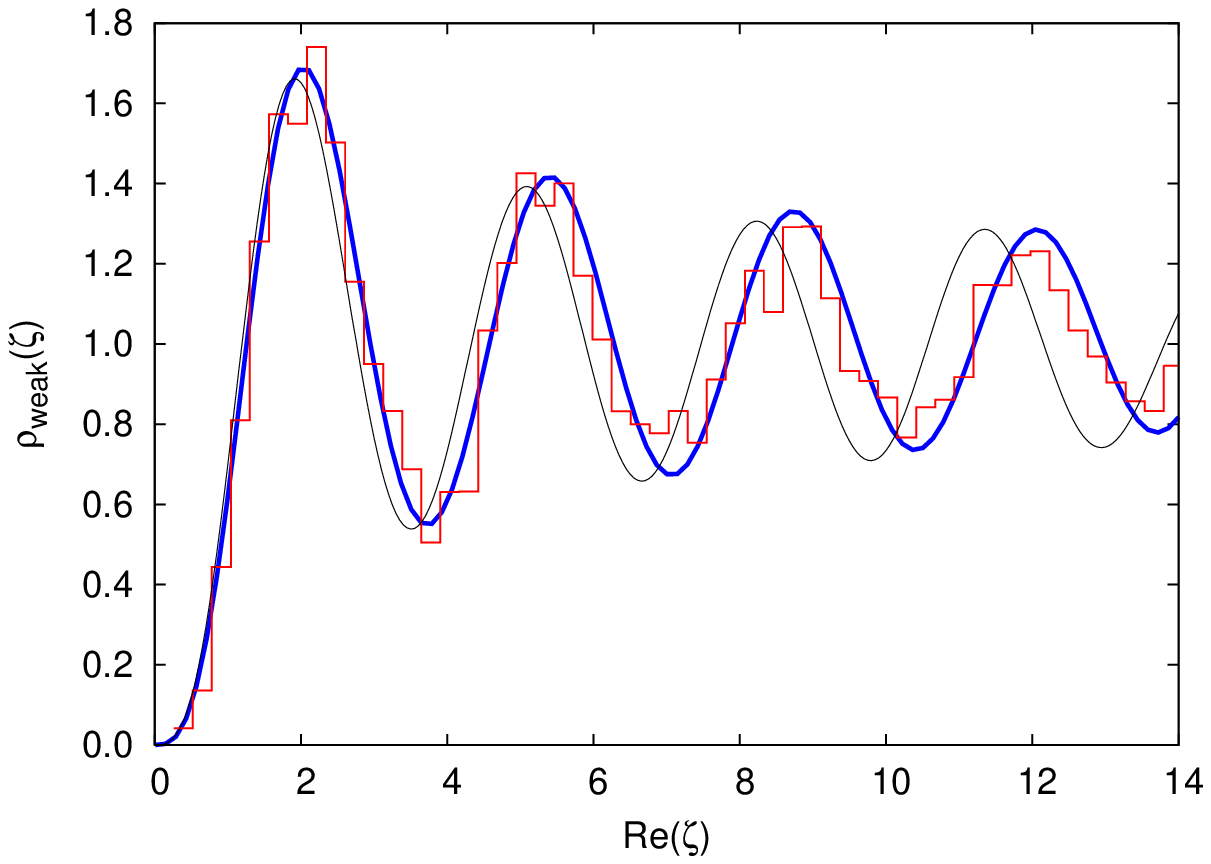}
\includegraphics[width=17pc]{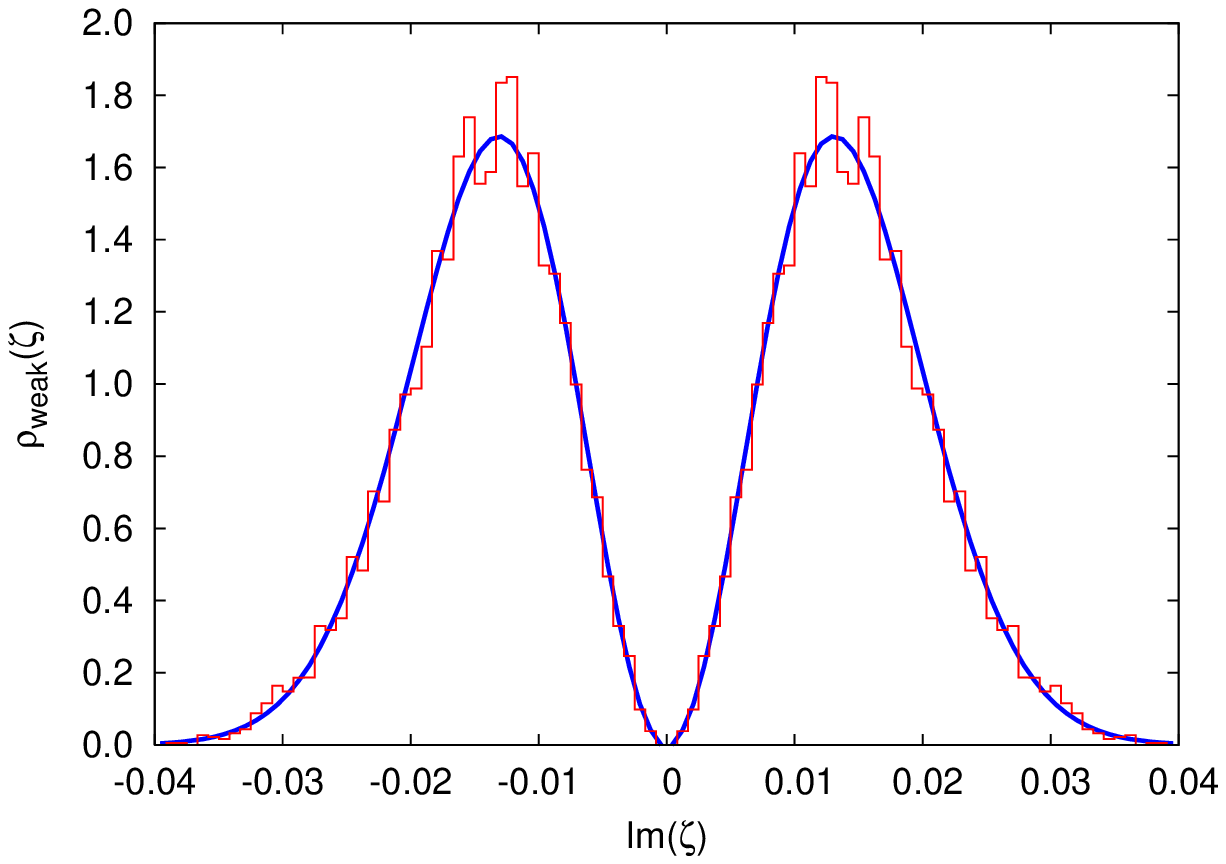}
\includegraphics[width=17pc]{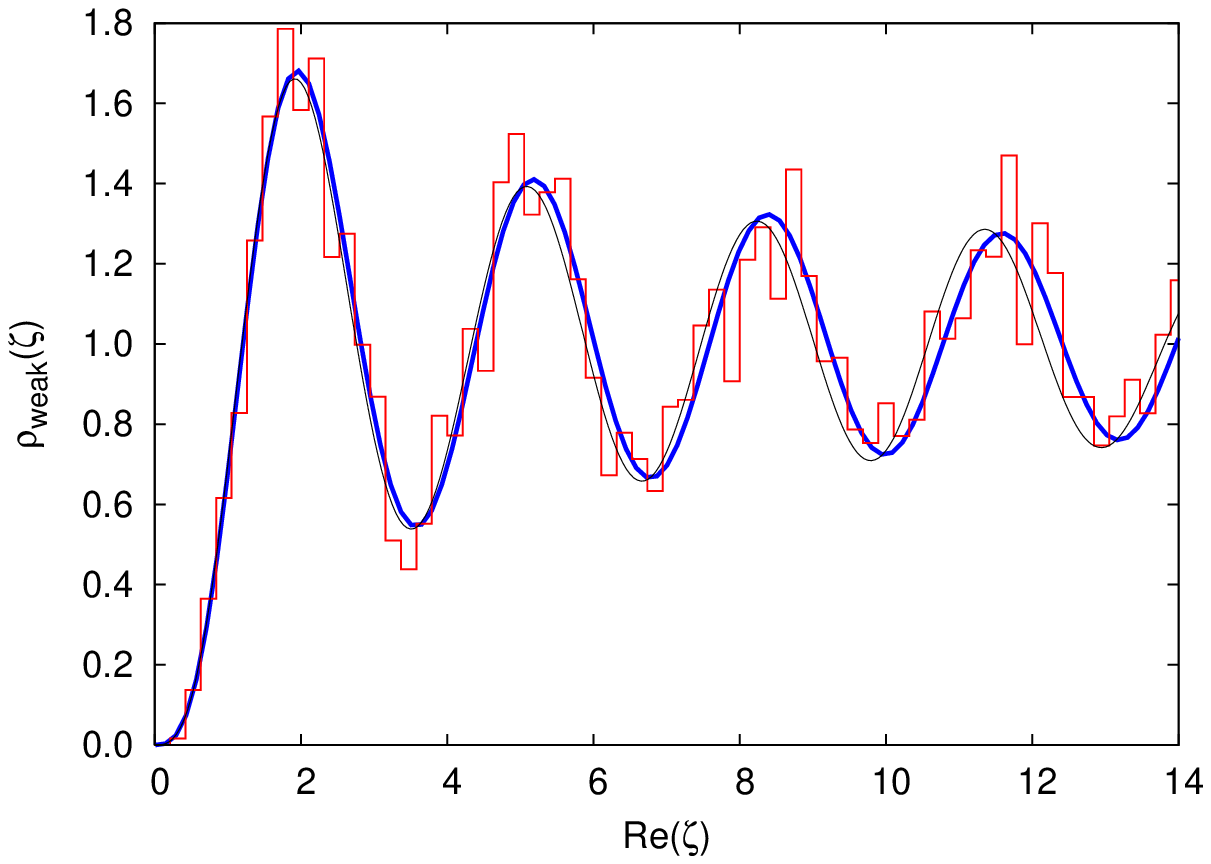}
\includegraphics[width=17pc]{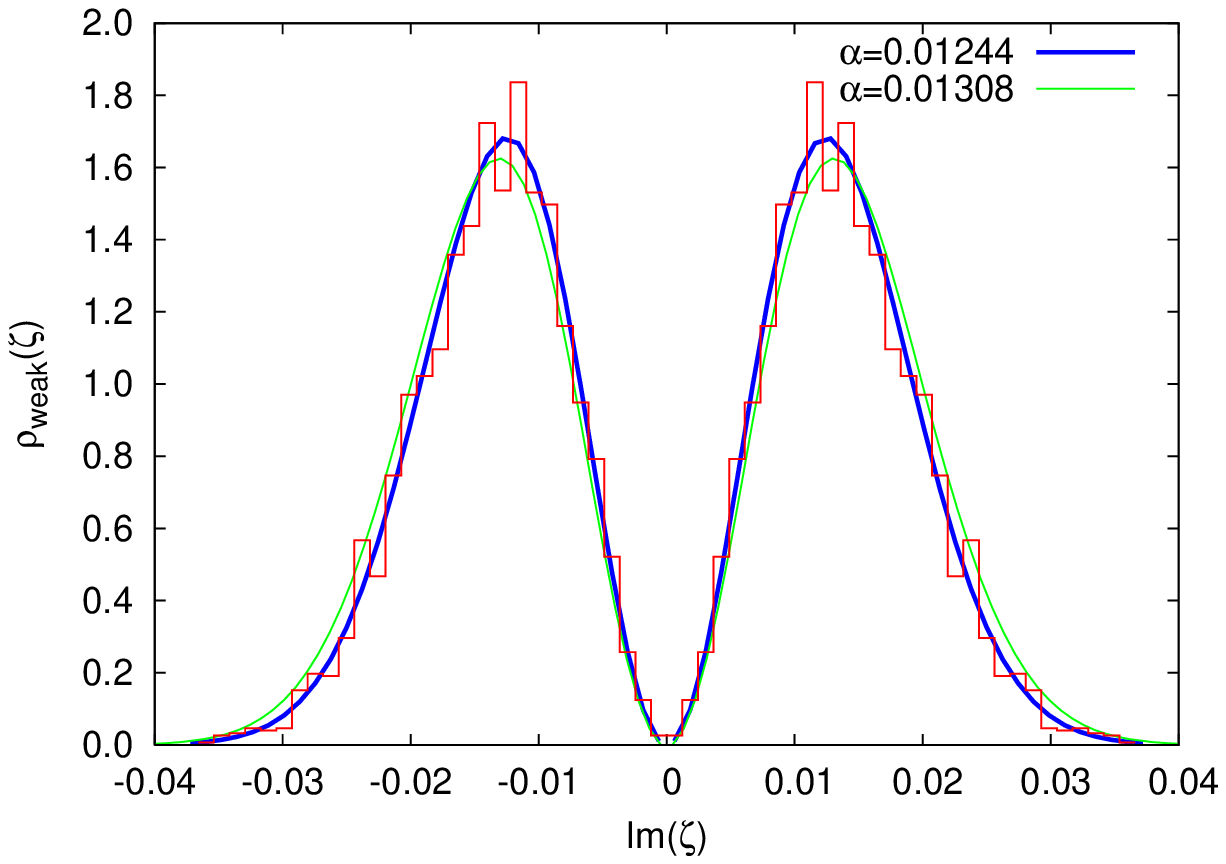}
\caption{The scaling $\mu^2V$ for dynamical fermions: 
$V=6^4$, $ma=0.07$ and $\mu=1\cdot10^{-3}$ $\Rightarrow$ $\alpha=0.013$,
  $\eta=33$ 
  (upper plots)  vs. $V=8^4$, $ma=0.07$ and  $\mu=5.625\cdot10^{-4}$
  $\Rightarrow$ $\alpha=0.012$, $\eta=102$  (lower plots) .}
\label{scaling}
\end{figure}
We can now test the scaling hypothesis of $\mu^2$ eq. (\ref{weak}).
For this purpose we have kept  $\mu^2V$ fixed for the 
two volumes $V=6^4$ and $8^4$: $\mu=1\cdot10^{-3}$ and
$\mu=5.625\cdot10^{-4}$, 
respectively. Since the level spacing $d$ also depends on the mass we have
kept $ma$ {\it fixed}, leading automatically to different $\eta$-values for
different volumes\footnote{For the comparison 
in \cite{ABLMP} the different $\eta\gg1$ 
were both close enough to quenched.}. 
The data (histograms) in Fig. \ref{scaling} right confirm
this scaling very well: the fitted $\alpha=0.013$ from $V=6^4$ (blue curve)
describes the $V=8^4$ data as well (green). 
For $V=8^4$ we also display the $\alpha=0.012$ (blue) obtained from an
independent fit, they agree within 5\%. 
In the cuts in 
Fig. \ref{scaling} left these small variations in $\alpha$ cannot be seen. 
If we were to compare to the quenched density a different fit value for  
$\alpha\sim 0.0185\ (6^4)$ and $\alpha\sim0.0175\ (8^4)$ would be obtained
instead, describing the right curves equally well. 
However, in the left plots the 
quenched MM curve (grey) deviates from the unquenched (blue) one.
In the upper $V=6^4$ plot the discrepancy from $\eta=33$ 
to quenched can be clearly seen in the data 
as we capture up to the $4th$ maximum, 
whereas in  $V=8^4$ keeping $ma$ fixed implies $\eta=102$, taking 
us back to almost quenched.
\begin{figure}[-h]
\centering
\includegraphics[width=17pc]{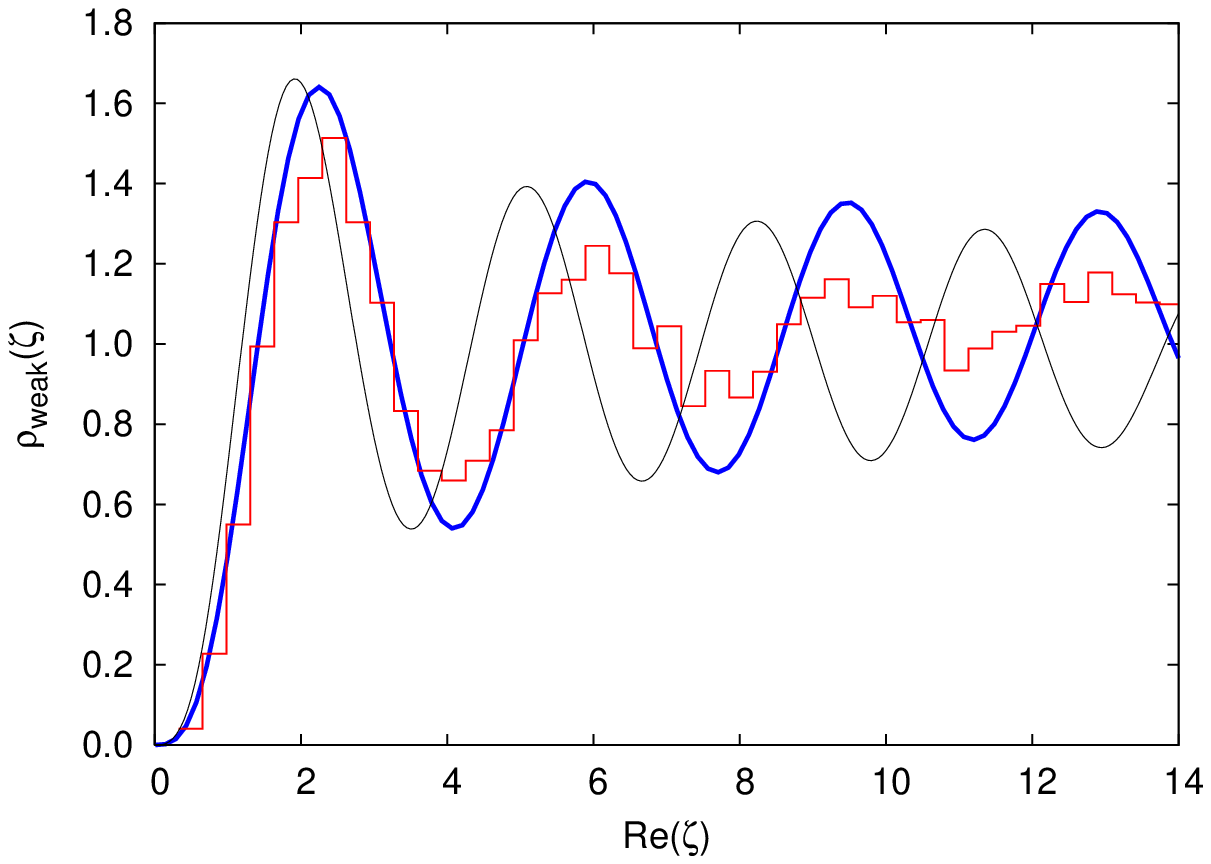}
\includegraphics[width=17pc]{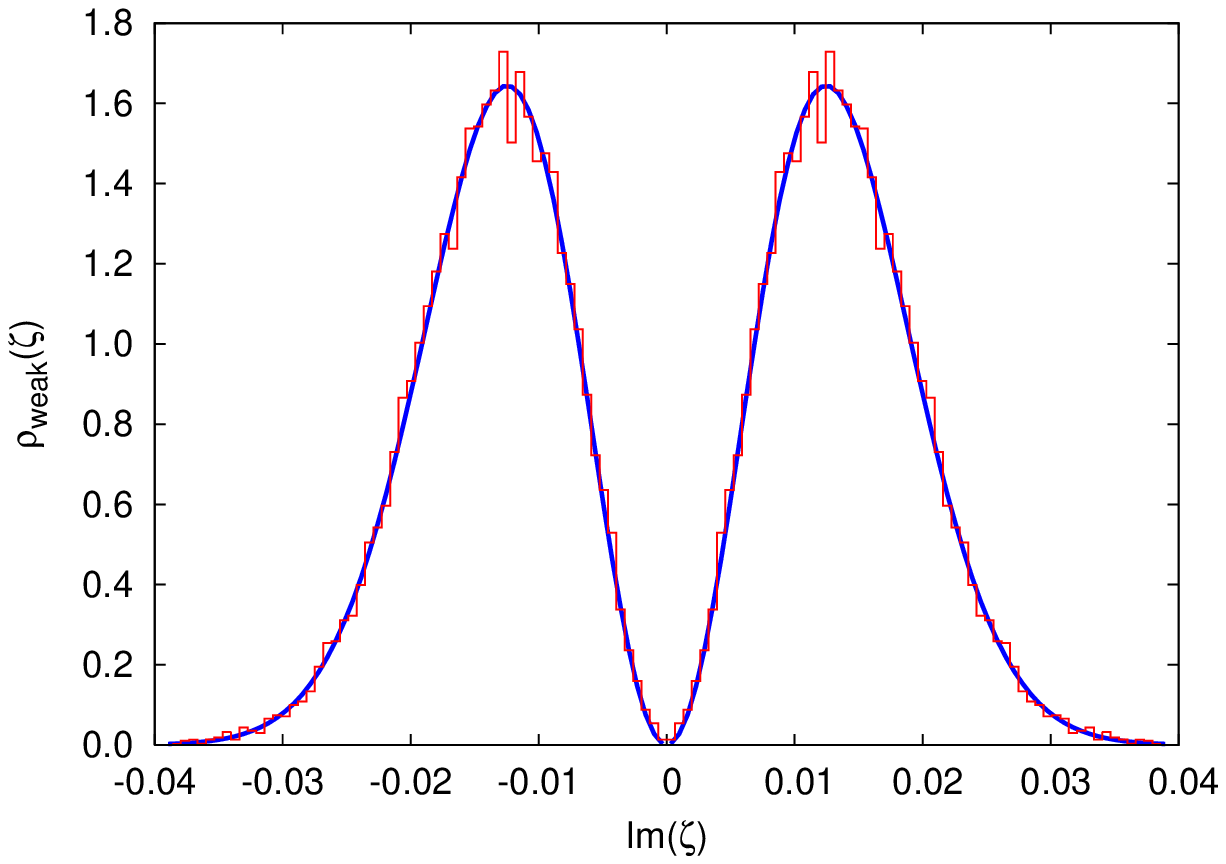}
\caption{Data (histograms) vs. MM (blue curve=unquenched, grey=quenched) 
for $V=6^4$ 
$ma=0.035$ and $\mu=1\cdot10^{-3}$ 
$\Rightarrow$ $\eta=14$ and $\alpha=0.012$.}
\label{small_eta}
\end{figure} 

In order to see the difference from quenched at smaller masses  
we compare to data corresponding to a rescaled mass $\eta=14$ (blue curve) in
Fig. \ref{small_eta} below. Although we can only resolve well 
the first 2-3 maxima
in the left picture, the mismatch with the quenched curve (grey) is evident.

To summarise we have shown that the MM correctly predicts complex $SU(2)$ 
Lattice data with $N_f=2$ dynamical staggered fermions at $\mu\neq0$, 
describing the effect of small quark masses. 
We have also confirmed the scaling of $\mu^2$ with the volume 
at weak non-Hermiticity from
unquenched Lattice data.

{\bf Acknowledgements:}
We thank Maria-Paola Lombardo, Harald Markum, Rainer Pullirsch 
and Tilo Wettig for
helpful discussions and previous collaboration. This work was supported by a
BRIEF Award No. 707 from Brunel University (G.A.) and by 
the Deutsche Forschungsgemeinschaft 
under grant No. JA483/17-3 (E.B.).


\begin{thebibliography}{99}

\bibitem{Jac3fold} J.J.M. Verbaarschot, Phys. Rev. Lett. {\bf 72} (1994) 2531.

\bibitem{HOV} M.A. Halasz, J.C. Osborn and J.J.M. Verbaarschot, 
Phys. Rev. {\bf D56} (1997) 7059.

\bibitem{A02} G. Akemann, 
Phys. Rev. Lett. {\bf 89} (2002) 072002; 
J. Phys. {\bf A}: Math. Gen. {\bf 36} (2003) 3363.

\bibitem{SV} K. Splittorff and J.J.M. Verbaarschot,
Nucl. Phys. {\bf B683} (2004) 467.

\bibitem{AFV} G. Akemann, Y.V. Fyodorov and G. Vernizzi,
 Nucl. Phys. {\bf B694} (2004) 59.

\bibitem{J} J.C. Osborn, 
Phys. Rev. Lett. {\bf 93} (2004) 222001; 
private communication.

\bibitem{AOSV} G. Akemann, J.C. Osborn, K. Splittorff and J.J.M. Verbaarschot
Nucl. Phys. {\bf B712} (2005) 287.

\bibitem{A05} G. Akemann, Nucl. Phys. {\bf B} to appear
  [\href{http://xxx.lanl.gov/abs/hep-th/0507156}{\tt hep-th/0507156}]. 

\bibitem{AW} G. Akemann and T. Wettig,
Phys. Rev. Lett. {\bf 92} (2004) 102002.

\bibitem{James}
J.C. Osborn,  {\it Eigenvalue correlations in quenched QCD at finite density},
these proceedings
PoS(LAT2005)200.	 

\bibitem{ABLMP} G. Akemann, E. Bittner, M.-P. Lombardo, H. Markum and 
R. Pullirsch,  Nucl. Phys. {\bf B140} Proc. Suppl. (2005) 568.

\bibitem{BLMP} 
E. Bittner, M.-P. Lombardo, H. Markum and
R. Pullirsch, Nucl. Phys. {\bf B106} Proc. Suppl. (2002) 
468.

\bibitem{MPW}  H. Markum, R. Pullirsch and T. Wettig,
        Phys. Rev. Lett. {\bf 83} (1999) 484.

\bibitem{Edouardo}  E. Follana, Nucl. Phys. {\bf B140} Proc. Suppl. (2005) 
141.

\bibitem{Karsch} F.  Karsch,
Prog. Theor. Phys. Suppl. {\bf 153} (2004) 106.

\bibitem{HKLM}
S. Hands, J.B. Kogut, M.-P. Lombardo, and S.E. Morrison, Nucl. Phys. B 558
(1999) 327.

\bibitem{BMW} M.E. Berbenni-Bitsch, S. Meyer and T. Wettig, 
Phys. Rev. {\bf D58} (1998) 071502.

\bibitem{AK} G. Akemann and E. Kanzieper, Phys. Rev.
Lett. {\bf 85} (2000) 1174.

\bibitem{Kogut}
J.~B.~Kogut, M.~A.~Stephanov, D.~Toublan, J.~J.~M.~Verbaarschot and
A.~Zhitnitsky, 
Nucl. Phys. {\bf B582} (2000) 477.



\end{thebibliography}
\end{document}